\documentclass[11pt,a4paper]{article}
\usepackage[english]{babel}
\usepackage[margin=1in]{geometry}
\usepackage{multicol}
\usepackage{background}
\usepackage[math]{blindtext}
 
 \usepackage{amsmath}
 \usepackage[cmintegrals]{newtxmath}

 \usepackage{fonttable}
 \usepackage{url}
 \usepackage{multicol}
 \usepackage{booktabs}
 \usepackage{lipsum}
 \usepackage{colortbl}
 \usepackage{multicol}
 \usepackage{multirow}
 \usepackage{kantlipsum} 
 \usepackage{mwe} 
 \usepackage{subcaption}
 \definecolor{mycolor}{rgb}{0.92,0.95,0.86}
 \definecolor{mycolor1}{RGB}{255, 204, 150}
 \definecolor{mycolor2}{RGB}{97, 48, 06}

\SetBgScale{4}
\SetBgColor{gray}
\SetBgAngle{90}
\SetBgContents{arXiv: some other text goes here}
\SetBgPosition{-1.5,-10}
\begin{document}
		\textbf{\textit{Submitted to a journal for publication.}}

 \vspace{1cm}
{\centering
 
{\bfseries\Large A survey on pseudonym changing strategies for \\ Vehicular Ad-Hoc Networks \bigskip}
 
Abdelwahab~Boualouache\textsuperscript{1} , Sidi-Mohammed~Senouci\textsuperscript{2} , and  Samira~Moussaoui\textsuperscript{1} \\

\vspace{0.5cm}
   
   {\itshape
\textsuperscript{1} Department of Computer Science, RIIMA laboratory, USTHB University, BP 32 El Alia Bab Ezzouar, Algiers, Algeria \\
\textsuperscript{2} DRIVE EA1859, Univ. Bourgogne Franche-Comt\'e, F58000, Nevers, France. \\

   }
}

\begin{abstract}
The initial phase of the deployment of Vehicular Ad-Hoc Networks (VANETs) has begun and many research challenges still need to be addressed. Location privacy continues to be in the top of these challenges. Indeed, both of academia and industry agreed to apply the pseudonym changing approach as a solution to protect the location privacy of VANETs' users. However, due to the pseudonyms linking attack, a simple changing of pseudonym shown to be inefficient to provide the required protection. For this reason, many pseudonym changing strategies have been suggested to provide an effective pseudonym changing. Unfortunately, the development of an effective pseudonym changing strategy for VANETs is still an open issue. In this paper, we present a comprehensive survey and classification of pseudonym changing strategies.  We then discuss and compare them with respect to some relevant criteria. Finally,  we highlight some current researches, and open issues and give some future directions. 
\bigskip
 
\textit{\textbf{Keywords:}}\noindent VANETs, security, location privacy, pseudonym changing.
 
\end{abstract}

\newpage
 
\section{Introduction}
\label{intro}

Over the last decade, Vehicular Ad-Hoc Networks (VANETs) have attracted a lot of interest from both the research community and the automotive industry due to their a huge impact on the future transportation systems (ITS) \cite{sommer}. This technology is primarily developed to enhance road safety and provide traffic efficiency. VANETs allow vehicles not only to communicate between them (V2V), but also with an installed  infrastructure (V2I), which enables a variety of interesting applications. These applications can be ranging from safety-related applications, such as collision warning and emergency reporting to non-safety applications like infotainment \cite{survey}. Safety-related applications are usually based on beaconing i.e. the process of periodically broadcasting safety messages. These latter are called Cooperative Awareness Messages (CAMs) in Europe and Basic Safety Messages (BSM) in US \cite{sommerivc} and they are broadcasted over the DSRC control channel (CCH) with a high frequency ranging from 1 to 10 Hz as suggested by standardization bodies such as IEEE, ETSI, and SAE \cite{doukha}. Safety messages include sensitive information about the current state of vehicles such as their identifiers, positions, and velocities. The encryption of these messages is not recommended since many VANETs' participants are concerned by them \cite{sevecomreport}. In addition, decrypting safety messages can add a latency in the processing of them, which may not meet with real-time requirements of safety-related applications \cite{realtime}. However, due to security threats such as false data injections, disseminated messages modifications, and reply attacks, safety messages must be authenticated.

The aim of safety messages is to make vehicles aware about their surrounding environment, which significantly improves road safety. For example, using these messages, vehicles can expect or detect dangerous situations that can cause serious damages on VANETs such as collisions and accidents. As a result, vehicles can then make decisions to prevent such bad consequences. However, although, safety messages are beneficial for road safety, they may also be exploited by adversaries for unauthorized location tracking of vehicles \cite{emara}. Indeed, due to the nature of the wireless medium, a passive adversary can easily eavesdrop all the broadcasted safety messages within its region of interest. It can then collect these safety messages and determine the locations visited by vehicles over time. The location tracking of vehicles could violate drivers privacy since one vehicle is usually associated only to one driver \cite{slotswap}. Therefore, knowing vehicle's position can lead to disclosure critical information about driver's life. For example, having information about of the frequency of driver's visits to a given hospital may raise doubts of the employer about the driver's health \cite{privaserv}. Furthermore, the driver's life can be put at risk if the adversary is a criminal. Protecting the location privacy is thus crucial because the lack of the protection may disturb the deployment of VANET technology.

Several privacy requirements for VANETs are identified in the literature (discussed in Subsection~\ref{sec:privacyrequirements}). The anonymity is one of the main privacy requirements. It ensures that safety messages are authenticated without attaching the real senders' identifiers. The anonymity is however contracted with the accountability security service, which aims to ensure that the authorities are always able to identify vehicles in case of a misbehaving behavior. Therefore, the privacy in VANETs must be conditional, where vehicles are anonymous to all VANETs' participants except the authorities, which must still track them. 

In order to meet these requirements, many anonymous authentication schemes have been proposed. These schemes can generally be divided into three categories \cite{reviewsecurity2015}: (i) the group-signature-based schemes (e.g. \cite{gsis}), (ii) pseudonymous authentication schemes (e.g. \cite{sevecom}), and finally (iii) hybrid schemes (e.g. \cite{calan}). However, both of academia and industry have adopted a pseudonymous authentication scheme to be implemented in the future deployment of VANETs \cite{motiv}. Indeed, the current security standards IEEE 1609.2 standard \cite{ieeesecuritystandard} and ETSI  102941-v1.1.1 \cite{etsi} are based on a traditional public infrastructure (PKI). The pseudonyms represent a set of certified public keys (anonymous certificates) stored in the vehicle's On-Board Unit (OBU) \cite{sevecom}. Instead of using one identifier (public key) all the time, a vehicle periodically changes its pseudonym to mitigate the tracking of its positions. Moreover, the change of pseudonym should be accompanied by the change of all the identifiers of communication stack layers such as the MAC and the IP addresses \cite{papadimi}. The pseudonym can then be seen as a fictive vehicle's identifier, where only the authorities can resolve it i.e. finding the relationship between a given pseudonym and the corresponding real identifier of the vehicle. The location privacy protection using the \textit{pseudonym changing approach} mainly depends on two factors: (i) the frequency of  pseudonyms changing i.e. the higher frequency of pseudonyms changing is, the more level of location privacy protection is. However, knowing that the pseudonyms are the identifiers that are used in the inter-vehiculaire communications, the changing of pseudonyms with a high frequency will certainly has negative effects on the communication performances. Indeed, Schoch and al. \cite{impact} demonstrated that a high level of packet loss is engendered when incorporating a geographic routing protocol with the changing of pseudonyms (frequency less than 30s), and (ii) the \textit{unlinkability} i.e. two pseudonyms belongs to the same vehicle should not be linked to each other.

Unfortunately, several conducted works to study the efficiency of the pseudonym changing approach (discussed in Subsection~\ref{sec:privacyeffiency}) demonstrated that a simple changing of pseudonym is ineffective to provide the required level of location privacy protection for the VANETs' users. This is due to the pseudonyms linking attack. Indeed, there exist two types of pseudonyms linking attack (presented in Subsection~\ref{sec:pseudoslinkingattacks}): the syntactic linking and the semantic linking. Many pseudonym changing strategies have been suggested to provide protection against this attack. The aim of a pseudonym changing strategy is to determine where, when, and how vehicles should change their pseudonyms to provide the unlinkablity between them. However, although, the variety of the existing the pseudonym changing strategies, there is no strategy suggested by standardization bodies to be applied until now \cite{bigdata}.

Several surveys have been conducted on the security and privacy in VANETs (e.g \cite{survey1, pseudonymapproach,survey2,survey3,survey4}). Petit et al. \cite{pseudonymapproach} presented a detailed survey on existing pseudonymous schemes in VANETs. The authors compared and classified the different existing solutions, and identified some open challenges among them the lack of an effective pseudonym changing strategy and the absence of a comparison between existing strategies. To complement these efforts, this paper presents a comprehensive survey of  existing pseudonym changing strategies for VANETs. The paper carefully analyzes these strategies to identify the strengths and weaknesses of each strategy. To the best of our knowledge, we are the first to propose such survey.  We hope that this survey will help to take a decision of which strategy should be applied in the future deployment of VANETs and potentially propose new strategies. 

The main contributions of this paper can be summarized as follows:
\begin{itemize}
	\item We survey and elaborate a taxonomy of pseudonyms changing strategies for VANETs.
	\item We discuss, analyze, and compare the presented strategies.
	\item We highlight some open challenges on the pseudonym changing strategy issue.
\end{itemize}

The rest of the paper is organized as follows. In Section 2, we present some necessary background information. A taxonomy pseudonym changing strategies is presented in Section 3. In Sections 4, we discuss, classify and compare the presented strategies. Some research challenges are given in Section 5. Finally, Section 6 concludes this survey. 

\section{Background }
\label{sec:background}

The purpose of this section is to give the reader the necessary background information to understand the research presented in this paper.

\subsection{Privacy requirements}

\label{sec:privacyrequirements}

The privacy is one of  the important human rights that should be protected \cite{hr}. However, the current technological development has threaten this right and reduced the control of users on their personnel information \cite{it-privacy}. For this reason, researchers have suggested a set of requirements to ensure the privacy protection of these technologies' users. Indeed, the following privacy requirements have been identified for VANETs \cite{requirements}:

\begin{itemize}
	
	\item \textbf{Minimum disclosure}: the amount of information revealed by a user should be limited to the necessary information to ensure VANETs' functionalities.  
	
	\item \textbf{Anonymity}: messages sent by a vehicle should be anonymous within a set of potential vehicles. This requirement contradicts the accountability, which is one of the main security requirement for VANETs. The accountability states that the authorities should be able to identify the origin of any sent message. For this reason, the anonymity should be conditional in VANETs.
	
	\item \textbf{Unlinkability}: two messages related to the same vehicle cannot be linked for longtime.
	
	\item \textbf{Perfect forward privacy}: resolution or revocation of one credential should not affect the unlinkability of any of the vehicle's other credentials.
	
\end{itemize}

\subsection{Studies on the efficiency of the pseudonym changing approach}
\label{sec:privacyeffiency}

Several studies have been conducted on the effectiveness of pseudonym changes. In the following, we discuss some existing studies. Butty\'{a}n et al. \cite{mixzone} studied the impact of the adversary power on the effectiveness of pseudonym changing approach. They found that if the adversary only controls the half of the road intersections then the probability of success tracking reaches 90\%. Wiedersheim et al. \cite{enoght} used an advanced tracking method called Multi-Hypothesis-Tracking (MHT) \cite{mth} incorporating with the Kalman filter \cite{kalman} to track vehicles. They claimed that a global passive adversary can effectively track the vehicles with accuracy of almost 100\%. In \cite{emara}, Emara et al. showed that the tracking can effectively be done even using a simple tracking method, called the Nearest Neighbor Probabilistic Data Association (NNPDA) \cite{nnpda}. In a recent study presented in Black Hat conference \cite{blackhat}, Petit et al. demonstrated through real-world experiments using ITS hardware that the location tracking of vehicles can easily be performed. Indeed, they found that the tracking success is achieved 40\% using only two sniffing stations, and 90\% if 8 sniffing stations are used.

The results of all these studies  confirmed that  vehicle's positions can be tracked even with the frequently changing of pseudonyms. This can be done using the pseudonyms linking attack. This attack is presented in the next subsection.

\subsection {Pseudonyms linking attack}
\label{sec:pseudoslinkingattacks}
Two types of pseudonyms linking attack have been identified by Butty\'an et al. in \cite{slow}. These types are described in the following subsections.

\subsubsection{Syntactic linking}
\begin{figure}[!ht]
	\begin{center}
		\includegraphics[width=8cm,height=4cm]{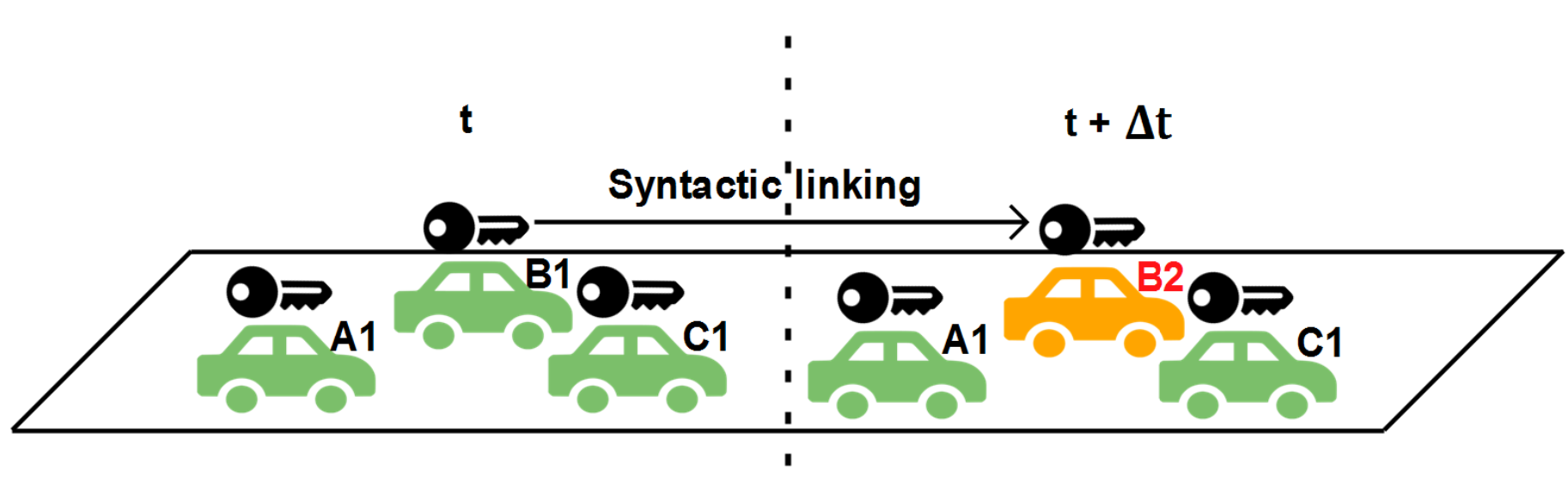}
	\end{center}
	\caption {The syntactic linking of pseudonyms}
	\label{fig:syntacticlinking}
\end{figure}

Figure~\ref{fig:syntacticlinking} illustrates the syntactic linking of pseudonyms case. If during {\itshape $\Delta$t} only one vehicle (\textit{B}) changes its pseudonym (from \textit{B1} to \textit{B2}) among the three vehicles that running on the road, the adversary can then easily link the pseudonyms \textit{B1} and \textit{B2}. The protection against this type of attack can be performed through using a mechanism to synchronize the changes of pseudonyms between vehicles.
\subsubsection{Semantic linking}

\begin{figure}[!ht]
	\begin{center}
		\includegraphics[width=8cm,height=4cm]{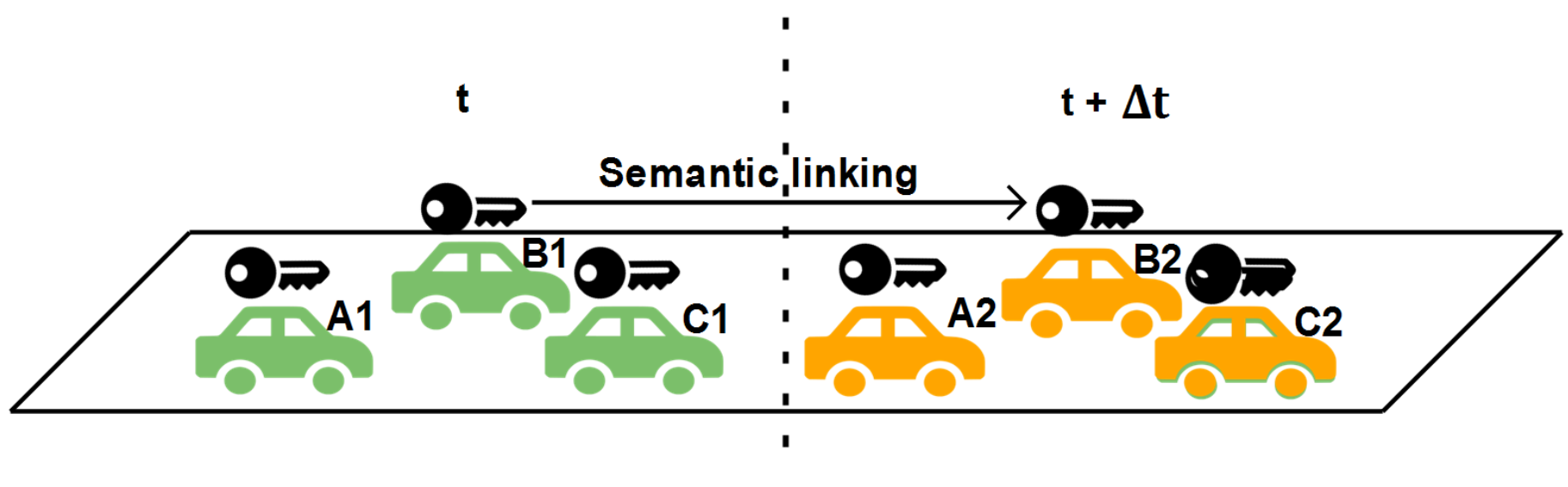}
	\end{center}
	\caption {The semantic linking of pseudonyms}
	\label{fig:semanticlinking}
\end{figure}

Figure~\ref{fig:semanticlinking} illustrates the semantic linking of pseudonyms. This type of attack is more powerful than the syntactic linking of pseudonyms because the adversary relies on the information included in safety messages to link the pseudonyms. For example, the adversary can predict the next position of the vehicle using a tracking method like \cite{enoght}\cite{emara}. Then, based on this prediction the adversary can link the pseudonyms \textit{B1} and \textit{B2} even if the three vehicles, illustrated in Figure 2, change their pseudonyms in the same time. The protection against this type of attack can only be done by preventing the adversary to get access to safety messages for some periods of time.

\subsection {Privacy metrics}
\label{sec:metrics}

A variety of metrics have been proposed to evaluate the level of the location privacy protection achieved by a pseudonym changing strategy. In \cite{privacymetrics}, Wagner et al. reviewed the different metrics used in this context. In the following, we present the most used metrics:

\begin{itemize}
	
	\item \textbf{Anonymity set size}: the anonymity set, denoted by AS, is defined as the set of vehicles that are indistinguishable from the target with the set including the target itself \cite{amoeba}. The size of the anonymity set $\mid$AS $\mid$  is then the number of vehicles that the anonymity set includes. The size of the anonymity set represents the level of the location privacy protection achieved, which should be grater than 1 in this case. However, this metric assumes that all vehicles of the anonymity set are equally likely to be the target. Therefore, as discussed in \cite{serjantov}, the knowledge of the adversary that makes some  vehicles more likely to be the target than the others cannot be described using this metric. For this reason, the entropy is suggested as a metric \cite{serjantov}.
	
	\item \textbf{Entropy of the anonymity set size}: the entropy is a concept coming from the information theory that expresses the uncertainty in a random variable. In contrast to the anonymity set size, the entropy of the anonymity set, denoted by H$_p$, allows expressing the adversary's knowledge about each vehicle of the anonymity set. The entropy is calculated using the following formula:
	
	\[ H_{p} =  - {\sum_{i=1}^{\mid AS \mid}} p_{i} \log_{2} p_{i} . \]
	
	Where p$_{i}$ refers to the probability of a vehicle i being the target. If all vehicles have a same probability to be the target i.e.  the probabilities are uniformly distributed over the anonymity set, the entropy then achieves its maximum value, denoted by H$_{max}$, which is given by:

	\[ \forall i : p_{i} = \frac{1}{\mid AS \mid},  H_{pmax}= - {\sum_{i=1}^{\mid AS \mid}} p_{i} \log_{2} p_{i} = \log_{2} \mid AS \mid. \]

	If we assume that the adversary has initially no knowledge about the vehicles of the anonymity set. Then, the information that can be obtained the adversary can be measured using the following difference: H$_{max}$ - H$_{p}$. \cite{entropy} proposed the degree of anonymity d, which a  normalized value of (H$_{max}$ - H$_{p}$) in the range [0,1]. The degree of anonymity is then  computed using the following formula:
	
	\[ d= 1-\frac{H_{max}-H}{H_{max}}= \frac{H}{H_{max}}  \]
	
	\item \textbf{Adversary's success rate}: the adversary's success rate is generally defined according to the proposed pseudonym changing strategy. It represents the ratio of vehicles that could still be tracked by the adversary after executing the strategy.
	
	\item \textbf{Maximum tracking time}: the maximum tracking time measures the maximum duration of time that the adversary stills linking the pseudonyms of vehicles. 
	
	\item \textbf{Statistics on pseudonym changes:} this can include information about changed pseudonyms such as their total number and the number of successful changes.
	
\end{itemize}

\subsection {Adversary model}

\label{sec:adversarymodel}
Due to the complexity of the VANET system, different attacks can be performed by different types of adversary. The potential types of adversary in VANETs have been extensively studied in the literature. \cite{raya2} identified the following types of adversary:  

\begin{itemize}
	\item \textbf{Global vs. Local}: compared to a local adversary, a global adversary has an overall coverage of the VANET. It can then eavesdrop every message diffused by any vehicle. 
	
	\item \textbf{Active vs. Passive}: an active adversary is more dangerous than a passive adversary since it can alter or inject messages, while a passive attacker can only eavesdrop messages.
	\item \textbf{Internal vs. External:} an internal adversary is an authenticated member of the VANETs system. An external adversary is considered as an intruder.
\end{itemize}

A location privacy adversary model aims to track the target vehicle by eavesdropping all communications of any vehicle within a region of interest. For this reason, researchers often consider the global passive adversary to study the location privacy in VANETs. \cite{annoucing} pointed out that due to the cost of eavesdropping the global coverage, is hard to be achieved. In \cite{blackhat}, the authors defined more realistic adversary called mid-sized adversary. The coverage of this adversary is larger than a local passive adversary and less than a global passive adversary. In other words, it can cover a limited number of areas without getting the full coverage. The authors also suggested to take into the account the tracking period to evaluate the power of this adversary. The tracking period represents time that the adversary try to link the pseudonyms of vehicles. Based on this latter parameter, the authors distinguished tree tracking types:

\begin{itemize}
	\item \textbf{Short-term tracking:} the adversary tries to track vehicles only for a couple of seconds.
	\item \textbf{Mid-term tracking:} the adversary tries to track vehicles  for a single trip, which can go from a couple of minutes to a couple of hours.
	
	\item \textbf{Long-term tracking:} in this case the period in which the adversary tries to track vehicles is extended to several days.
\end{itemize}

\begin{figure*}[!ht]
	\begin{center}
		\includegraphics[scale=0.6]{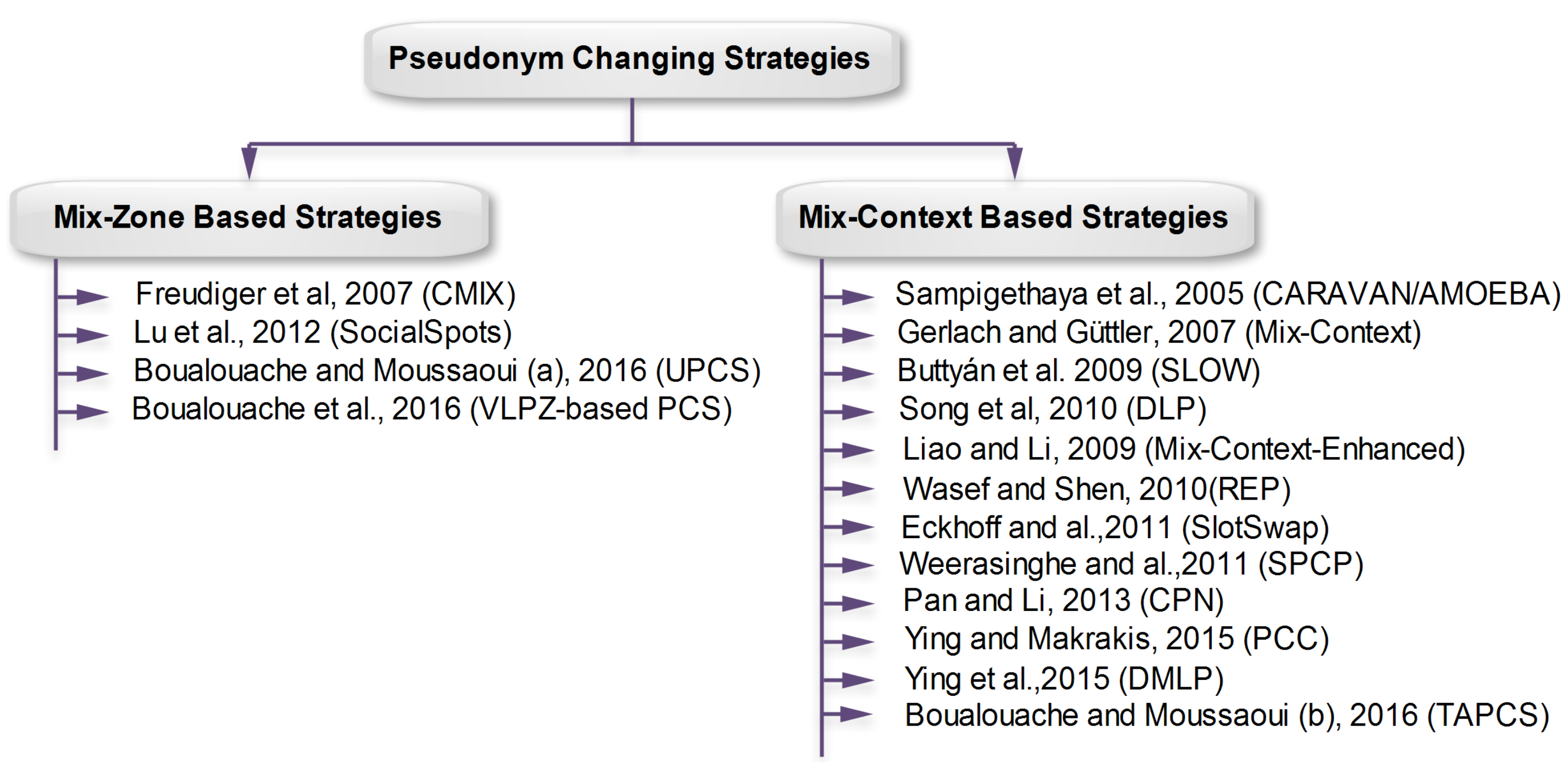}
	\end{center}
	\caption {Pseudonym changing strategies taxonomy.}
	\label{fig:taxonomy}
\end{figure*} 

\section{Pseudonym changing strategies: a taxonomy}
\label{sec:taxonomy}

The main purpose of a pseudonym changing strategy is to determine where and when a vehicle should change its pseudonyms  to achieve the unlinkablity between them. Many pseudonym changing strategies have been proposed to provide protection against the pseudonyms linking attack. In Figure~\ref{fig:taxonomy}, we propose a taxonomy of existing pseudonym changing strategies for vehicular ad-hoc networks. We divide these strategies into two categories: (i) mix-zone-based strategies, and (ii) mix-context based strategies.

\subsection{Mix-Zone-based strategies}

In mix-zone-based strategies, vehicles change their pseudonyms on predefined road areas, called mix zones. The concept is first proposed by Beresford and Stajano in the context of pervasive computing \cite{mixidea}. The authors in \cite{mixzone}, studied the location privacy protection against a limited adversary model that can only control a limited number of places of the vehicular area. They then considered the regions that are not controlled by the adversary as mix zones. 

\begin{figure}[!hb]
	\begin{center}
		\includegraphics[width=7cm,height=5cm]{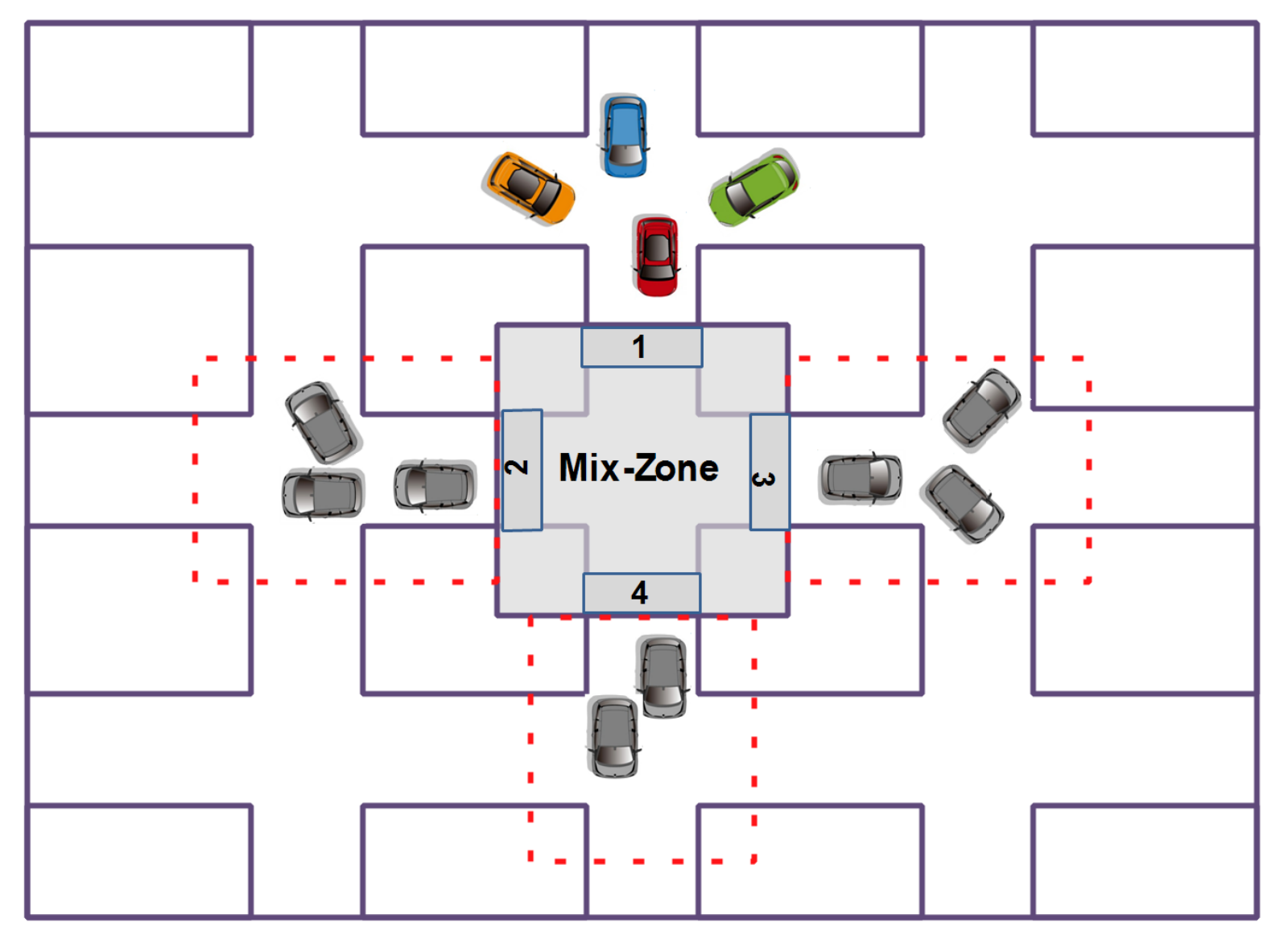}
	\end{center}
	\caption {Mix-zone concept}
	\label{fig:mixzone}
\end{figure} 

Figure~\ref{fig:mixzone} illustrates a mix zone installed at a road intersection. If a vehicle enters the zone from the port 1, changes its pseudonym inside the mix zone, and after that exits it from one the port 2,3,4 the adversary could confused due of this.

The first implementation of the mix-zone concept proposed by Freudiger et al in \cite{cmix}. The authors proposed a protocol for creating CMIX (Cryptographic MIX) zones. A CMIX zone is a road area where safety messages are encrypted. The authors suggested placing these mix zones at road intersections. Vehicles change their pseudonyms inside a CMIX zone and use a shared key distributed by a RSU to encrypt their safety messages. Each intersection is equipped by a RSU, which periodically broadcasts a notification to inform vehicles the existence of a CMIX zone. When a vehicle receives such notification, it sends a message to the RSU to request the encryption key. As soon as the RSU receives the request, it provides the encryption key to the vehicle and waits for an acknowledgment from it. If the vehicle well received the key, it sends the acknowledgment to the RSU, and starts the encryption of its safety messages using this key. It must also change its pseudonym within the CMIX zone. The authors also pointed out the problem of vehicles entering to a CMIX zone that did not receive a notification for the RSU  yet. Indeed, these vehicles are not able to decrypt the safety messages coming from the CMIX zone. For this reason, they proposed that the encryption key should be forwarded from the vehicles that are already inside the CMIX zone to the vehicles entering to it.

Besides of this, several works have be been conducted to propose the optimal deployment of CMIX zones over the road's intersections (e.g. \cite{optimalcmix1}\cite{optimalcmix2}\cite{optimalcmix3}\cite{optimalcmix4}\cite{optimalcmix5}). In these works some optimization techniques such as the multi-objective optimization,  heuristics, and the game-theoretic approach are used to find the optimal deployment of CMIX zones to achieve high levels of location privacy protection.  

Lu et al. \cite{socialconf,social} suggested to change the pseudonym at Social Spots, which are simply public places areas such as signalized intersections, when traffic light turns to red, and parking lots near a shopping mall. Two simple pseudonyms changing strategies are then proposed in these papers: (i) all vehicles stopped in front of the red traffic light at signalized intersections,  change their pseudonym together when the traffic light turns green, and (ii) each vehicle stopped at a free parking lot near a shopping mall, changes its pseudonym just before leaving the parking lot.

Boualouache and Moussaoui (a) \cite{s2si, upcs} proposed a strategy adapted for the urban environment. This strategy is based on the creation silent mix zones at signalized intersections only while the traffic light is red. The authors suggested that vehicles can either change or exchange their pseudonyms inside these silent mix zones. The signalized intersection is then equipped by a RSU, which starts broadcasting notifications only if the traffic light turns to red. Each notification includes the position of the beginning of the silent mix zone, denoted by P$_{sm}$. If a vehicle receives such notification, it compares its position with P$_{sm}$. If a vehicle  finds itself inside the silent mix zone, it immediately stops broadcasting safety messages. Two techniques can be used inside the silent mix zone. Indeed, the vehicle can either simply change its pseudonym or exchange it with other vehicle. The exchange of pseudonyms is performed using the swapping protocol, where each time, the RSU chooses randomly two vehicles to exchange their pseudonyms and informs the CA about each performed exchange to keep the accountability. All the messages used in the exchanging are encrypted. 

Boualouache et al. \cite{vlpz} proposed a pseudonym changing strategy based on a designed roadside infrastructure, called  the Vehicular Location Privacy Zone (VLPZ). The design of a VLPZ is similar to existing roadside infrastructures such as gas stations, electric vehicles charging stations, and toll booths.  A basic VLPZ consists of two points: (i) one entry point called \textit{the router}, and (ii) one exit point called the \textit{aggregator}; and  a limited number of lanes {\itshape l} where {\itshape l} $>$ 1. Vehicles arrive to a VLPZ, one after another, on one lane. When a vehicle reaches \textit{the router}, it stops broadcasting safety messages and heads for a VLPZ's lane randomly and privately assigned by \textit{the router}. Vehicles can reside inside a VLPZ for a random period of time. For example, if a VLPZ is deployed in a gaz station, this period is the time taken by the driver to fill the fuel tank of its vehicle. Vehicles must change their pseudonyms before they exit the VLPZ through the aggregator. However, the exit order is different from the entering order since the residency periods of vehicles are random.

\subsection{Mix-Context-based Strategies}

In contrast to  mix-zone-based strategies, in mix-context-based strategies, each vehicle independently determines where and when to change its pseudonym. This concept is first introduced in the context of wireless networks in general \cite{swing}. The concept is called a user-centric approach and suggests that mobiles users should be able to control their location privacy protection.  \cite{mixcontext} adapted this concept for the VANETs. Indeed, the authors defined the mix-context as any situation or opportunity helps the vehicle to increase its location privacy protection using the changing of pseudonym mechanism. Precisely, the mix-context is defined as any situation or opportunity to synchronize the pseudonyms changes between vehicles. The vehicle then changes its pseudonym only if the mix-context is found. 

\begin{figure}[!h]
	\begin{center}
		\includegraphics[width=9cm,height=10cm]{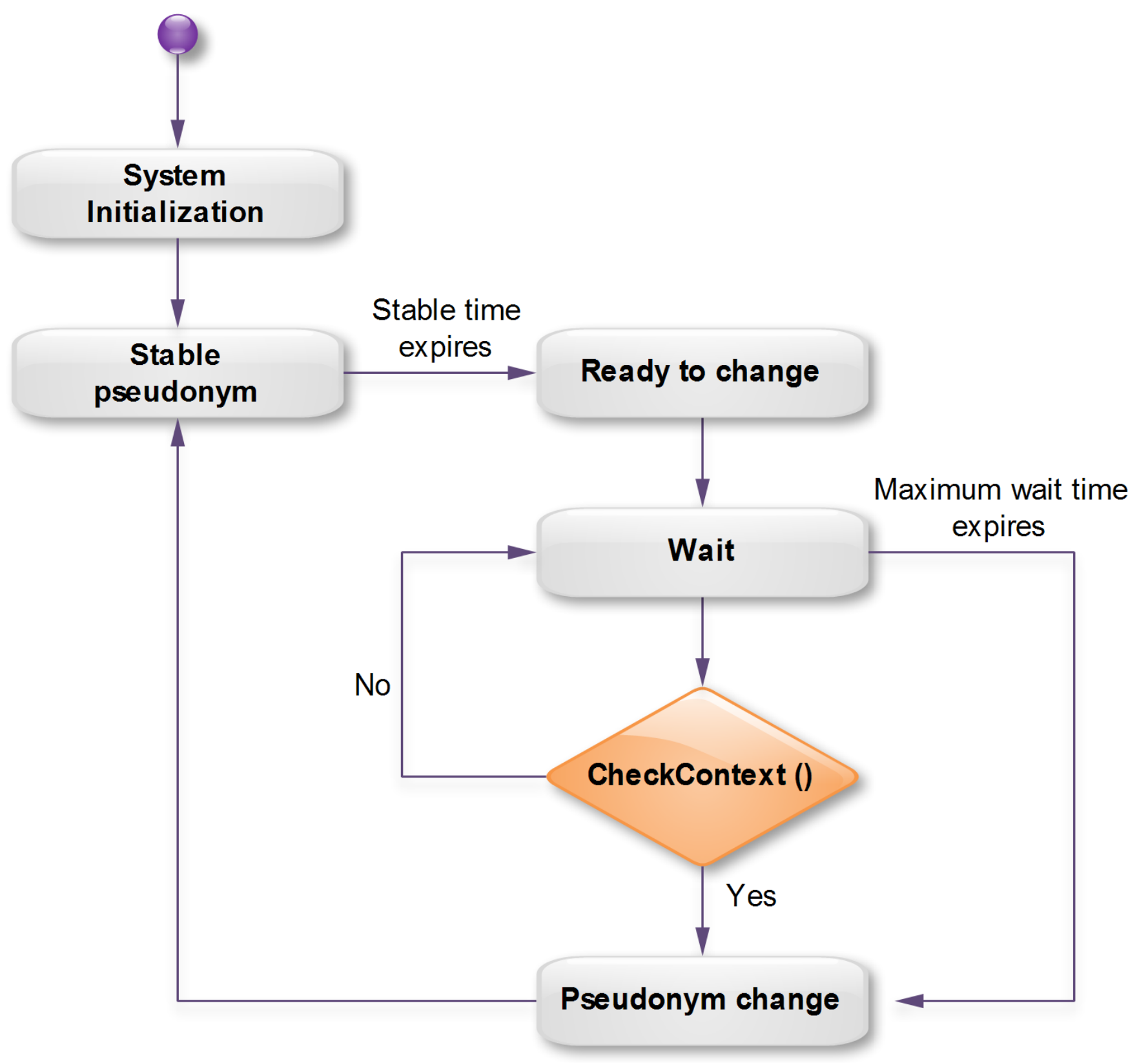}
	\end{center}
	\caption {A general state diagram of a mix-context-based strategy \cite{mixcontext} \cite{mixcontextenhanced}.}
	\label{fig:mixcontext}
\end{figure}

Figure~\ref{fig:mixcontext} illustrates a general state diagram of a mix-context based strategy. The vehicle is initially equipped by a set of pseudonyms. Each pseudonym is used for a limited period of time, which is called the stable time. The stable time  should be greater than a certain threshold to not affect safety-related applications \cite{impact} as discussed in the introduction. After the expiration of the stable time, the vehicle then moves to \textit{ready to change} state. It initializes a timer and starts looking for a mix context. If the mix context is found, the vehicle immediately changes its pseudonym. However, if the mix-context is not found after a certain threshold of time determined by the system designer, the vehicle is forced to change its pseudonym.


Gerlach and G\"uttler \cite{mixcontext} considered the direction and the number of vehicles in the communication range as mix context parameters. The vehicle changes its pseudonym only if it detects k neighboring vehicles at a distance smaller than the minimal distance and have a similar direction with it within its communication range. The minimal distance is considered equal to 4.25 meters in their simulations. Song et al. \cite{densityzone1,densityzone2} proposed the Density-based Location Privacy (DLP) strategy, which is based on a zone known as K-density zone to change the pseudonym. The vehicle changes its pseudonym within the K-density zone, if (k-1) neighbors are found over the communication range.  

Liao and Li \cite{mixcontextenhanced} proposed an improvement of \cite{mixcontext}. The authors proposed to add the speed, the distance between vehicles and the road segment in the mix context. They also proposed to insert a bit (flag) in the safety message, which indicates the willingness of a vehicle to change its pseudonym. The vehicle then changes its pseudonym if it found k neighboring vehicles have similar status to itself and whose flags equal to 1. The purpose of these improvements is to increase the probability that many vehicles simultaneously change their pseudonyms.

Pan and Li \cite{analysis,Cooperative} proposed  Cooperative  Pseudonym strategy (CPN). They first assume that the safety messages broadcasts are synchronized to be sent at the same time slot using GPS clocks. A vehicle \textit{v} changes its pseudonym if it found at least k neighboring vehicles ready to change their pseudonyms or one of its neighboring vehicles has at least k neighboring vehicles ready to change their pseudonyms. In other words, on the one hand, the vehicle \textit{v} cooperates with its neighbors if one of them has k or more neighbors ready to change their pseudonyms. On the other hand, if \textit{v} has k neighbors ready to change their pseudonyms, they will cooperate with it, if none of them has k neighbors ready to change their pseudonyms.

Ying and Makrakis \cite{condidates} proposed a pseudonym changing strategy based on a Candidate-location-list, called PCC. As \cite{Cooperative}, the authors assumed that the safety messages are synchronized to be broadcasted at the same time slot and in a multi-hops way. In addition, vehicles can reuse their pseudonyms more than one time. The authors proposed to insert new five fields into each broadcasted safety message: (i) \textit{sa1}: the address of the vehicle that generates the message, (ii) \textit{sa2}: the address of the vehicle that rebroadcasts the message, (iii) \textit{hop}: the number of hops between the source and the receiver, which is initially set at 0 and incremented each time the safety message is rebroadcasted by a vehicle, (iv) \textit{MaxLiveTime}: the lifetime of the message, and finally (v) \textit{ChStot}: that indicates on which slot the vehicle should change its pseudonym. These additional information aim to allow a vehicle to track the time slots on which other  vehicles change their pseudonyms. Each vehicle stores some information about the received safety messages for a limited storage time (\textit{MaxLiveTime}) in a Candidate Location List (CCL). Each entry of the CCL contains the following fields: \textit{ID}, \textit{hop}, \textit{ChSlot}, \textit{MaxLiveTime}, \textit{Position}, and \textit{Timestamp}, where \textit{ID} is the current pseudonym used by the vehicle, \textit{Position} is the current position of the vehicle, and \textit{Timestamp} is the time when the entry is inserted ou updated in the CCL. If a vehicle receives a safety message, it first checks if the value of \textit{hope} is not greater than \textit{MaxHopThreshold}. If this is the case, the vehicle then inserts an entry for this message in its CCL, increases the value of \textit{hop} by 1, and rebroadcasts it again. In addition, if the vehicle receives a safety message with a value of \textit{hope} less than the value of \textit{hope} saved in its CCL, it then updates the corresponding entry.

The PCC strategy is then executed as follows. The vehicle initially sets the value \textit{ChStot} to -1 and based on the prediction on the movement of vehicles, it periodically calculates the distance between itself and each vehicle saved in its CCL. It then chooses at least \textit{k} vehicles that have minimum distances from it. h$_{max}$ is defined as the maximum value of \textit{hop} of these vehicles. If a vehicle's pseudonym expires at time slot i, the vehicle should then broadcast a safety message with \textit{ChSlot} equals to i+1 at time slot i-h$_{max}$. If a vehicle receives such message, it rebroadcasts it, changes its pseudonym at time slot i+1. However, if during the lifetime of the current pseudonym, the vehicle receives several safety messages with different values of \textit{ChSlot}, it changes its pseudonym at the minimum value of \textit{ChSlot} and does not change it again until its expiration.

Sampigethaya and al. \cite{amoeba, caravan} adopted the use of the random silence periods concept \cite{silence}. The authors proposed that vehicles turn their radio transmitters off (stop sending safety messages) for a limited random period of time each time when are going to change their pseudonyms. The aim of this random silence period is to try to confuse the adversary. Indeed, If at same time two vehicles at least turn their radio transmitters off for a random period of time and change their pseudonyms during this period , the adversary could then be confused \cite{caps}\cite{poster}.

Butty\'an et al. \cite{slow} proposed a simple  pseudonym changing strategy, called SLOW (Silence at low speeds). This strategy uses the radio silence concept. Indeed, the authors suggested that vehicles turn their radio transmissions off when their speeds are less than 30Km/h and change their pseudonyms during the radio silence.
\nocite{reputationbased}
Ying et al. \cite{dynamicmix,motivationbased} proposed a pseudonym changing strategy, called DMLP (Dynamic Mix-Zone for Location Privacy in Vehicular Networks). This strategy aims to dynamically create CMIX zones i.e. the vehicle establishes a CMIX zone where and when it is needed. The authors introduced new entities in the VANET system architecture, called Control Servers (CSs). CSs are responsible to control and coordinate the change of pseudonyms processes between vehicles. Each CS is connected to a set of RSUs that cover a certain area. Each pseudonym is used by a vehicle for $\Delta t$. If a vehicle \textit{v} wants to create a DMLP, it should send a Request New Pseudonym (RNP) message to CS close to the expiration of its current pseudonym. The vehicle \textit{v} broadcasts the last message with the old pseudonym at \textit{t} and waits for the period of time ($\tau$) to change its pseudonym before begins broadcasting safety messages. $\tau$ is the period of time that takes the vehicle to create a dynamic mix zone. The dynamic mix zone is created as follows. If the CS receives a RNP message, it first determines the length (\textit{l}) of the zone that will be created. After that it sends a COMMAND message to the relevant RSUs that exist in the zone. As soon as a RSU, receives such message, it immediately rebroadcasts it to the vehicles within its communication range. If a vehicle receives a COMMAND message, it starts encrypting safety messages, changes its pseudonym and sends a RNP message to the CS. The concerned vehicles are still encrypting safety messages for T$_{EP}$, which is T$_{EP}$  $\leq$ $\Delta t$.

Wasef and Shen \cite{rep} suggested random encryption periods (REPs).  When the vehicle decides to change its pseudonym, it sends a request to its neighbors for starting a REP. During a REP, the safety messages are encrypted using a shared group key.  A REP is considered successful if at least one of the neighbors also changes its pseudonym and its speed or its direction. Eckhoff et al. \cite{sswapp,slotswap} suggested equipping vehicles with a limited number of pseudonyms. Each pseudonym is used for a specific short period of week.  The authors pointed out that pseudonyms can be linked, because always a same pseudonym is used during a same period. For this reason, they proposed using the exchanging of pseudonyms technique, which is inspired by \cite{swing}. The vehicles can then exchange their pseudonyms through encrypted channels. 

Weerasinghe et al. \cite{spcp} proposed a strategy called SPCP (pseudonym changing protocol). In SPCP, vehicles are self-organized as groups. Each group is managed by a group leader, which randomly decides when to change the pseudonyms of group's members. All members are then informed about the time when the pseudonym changing process will occur. During this process each member the group identifier as a temporal pseudonym and each vehicle quits the group should also changes its pseudonym. 

Boualouache and Moussaoui (b) \cite{tapcs} proposed a strategy, called TAPCS (Traffic-Aware Pseudonym Changing Strategy). In TAPCS, vehicles continuously monitor the road traffic's status to find a good place where the silent mix zone (SM) can be created. This strategy is mainly based on a privacy-preserving traffic congestion detection protocol that acts as a trigger of the strategy and consists of fives phases. (i) Traffic Congestion Detection Phase: In this phase, every vehicle continuously monitors its speed. If its speed is still lower than a certain threshold for a certain period of time, the vehicle reports a potential traffic congestion and broadcasts a congestion message. The detection of the traffic congestion is confirmed only if the vehicle receives a specific number of traffic congestion confirmations from the surrounding vehicles. (ii) TAPC strategy's Initiator Election Phase: After detecting the traffic congestion, a first initiator of the strategy will be elected. If the vehicle did not receive any initiation message from one of the surrounding vehicles and confirms the existence of traffic congestion, it stops broadcasting safety messages for a random time, changes its pseudonym, broadcasts an initiation message, and waits for a certain delay time. During this time, it stores each received initiation message. In the end of this time, the vehicle checks if its position is the minimum among the received positions. If so, the vehicle assigns itself as an initiator for the strategy.  (iii) Silent Mix Zone Creation Phase:  Just after its election, the initiator stops broadcasting safety messages, changes its pseudonym and starts broadcasting notifications, which contain the position and the direction of the initiator and the threshold of speed below of which vehicles stop broadcasting safety messages. If a vehicle receives such notification, it checks if it is situated in the same direction behind the initiator and its speed is lower than the speed threshold included in the notification, if so the vehicle then stops broadcasting safety messages and changes its pseudonym during this time. (iv) Silent Mix Zone Extension Phase: The SM zone is extended and a new initiator is elected before the SM zone gets filled. Indeed, each vehicle inside the SM zone calculates the distance between itself and the initiator. If the distance roughly equals to a fixed value the vehicle executes the initiator election phase as described in phase (ii). If the new initiator is elected, the previous one stops broadcasting notifications as soon as receives the first notification from the new initiator. and finally (v) End of Traffic Congestion Detection Phase : The extension of the SM zone process continues until the end of the traffic congestion; which is detected by the first initiator. Indeed, when its speed is still higher than the speed threshold for a certain period of time, it then broadcasts the end of congestion message. This message will be rebroadcasted by each elected initiator. If a vehicle receives this message, it restarts broadcasting safety messages again.

\subsection{Summary}
\label{sec: summary}
In this section, we provide a summary to get an overview of the presented pseudonym changing strategies. Table~\ref{tab:summary} summarizes characteristics and the methods used to evaluate the presented strategies. The parameters of this summary are presented as follows:

\begin{itemize}
	
	\item \textbf{Category}: indicates to which category the strategy belongs. 
	\item \textbf{Mode}: indicates whether the strategy needs the infrastructure to work or not.
	\item \textbf{Radio Silence}: indicates whether the strategy uses the radio silence or not.
	\item \textbf{Encryp}: indicates whether the strategy uses the encryption or not.
	\item \textbf{Pseudo Exchang}:  indicates whether the strategy uses the exchanging of pseudonyms or not.
	\item \textbf{Evaluation Metric}: indicates which metric is used to evaluate the level of the location privacy protection achieved by the strategy. These metrics are described in Subsection~\ref{sec:metrics} and abbreviated as follows:
	\begin{itemize}
		\item ASS: Anonymity Set Size.
		\item ASR: Adversary's Success Rate.
		\item MTT: Maximum Tracking Time.
		\item DA: Degree of Anonymity.
		\item Statistics: Statistics on Pseudonym Changes.
	\end{itemize}
	\item \textbf{Evaluation Method}: indicates which method is used to evaluate the level of the location privacy protection achieved by the strategy.

\end{itemize}

\begin{table*}[!ht]
	\centering
	\resizebox{\textwidth}{!}{
		\begin{tabular}{l l l l l l l l l l l}
			
			\hline 
			\toprule
			\multirow{2}{2.5 cm}{{\textbf{Strategy}}} & 
			\textbf{Category} & 
			\textbf{Mode}     & 
			\textbf{Radio}     &
			\textbf{Encryp} & 
			\textbf{Pseudo} & 
			\textbf{Evaluation } &
			\textbf{Evaluation} 
			
			\\ 	&  &  & \textbf{Silence} & &  \textbf{Exchang}   &  \textbf{Metric} & \textbf{Method}  \\
			
			\hline 
			\toprule
			
			\begin{tabular}[c]{@{}l@{}}CMIX \cite{cmix} \\ \end{tabular}& \begin{tabular}[c]{@{}l@{}}Mix-zone- \\ based \end{tabular} & \begin{tabular}[c]{@{}l@{}}Infrastructure-based \end{tabular} & No & Yes & No& \begin{tabular}[c]{@{}l@{}}Entropy \\ASR \end{tabular}& Simulation\\
			
			\hline
			\begin{tabular}[c]{@{}l@{}}CARAVAN/A-\\MOEBA \cite{amoeba,caravan}  \end{tabular}
			&  \begin{tabular}[c]{@{}l@{}}Mix-cont\\ext-based \end{tabular} & Infrastructureless & Yes & No & No  & \begin{tabular}[c]{@{}l@{}}Entropy\\MTT \end{tabular} & \begin{tabular}[c]{@{}l@{}} Analy-model \\ Simulation  \end{tabular}  \\
			
			\hline
			\begin{tabular}[c]{@{}l@{}} Mix-Context \cite{mixcontext}\\ \end{tabular}& \begin{tabular}[c]{@{}l@{}}Mix-cont-\\ext-based \end{tabular} & Infrastructureless & No & No & No & MTT & Simulation  \\

			\hline
			\begin{tabular}[c]{@{}l@{}} SLOW \cite{slow}\\ \end{tabular}& \begin{tabular}[c]{@{}l@{}}Mix-cont\\ ext-based \end{tabular} & Infrastructureless  & Yes & No & No & ASR & Simulation  \\
			
			\hline
			\begin{tabular}[c]{@{}l@{}} DLP \cite{densityzone1,densityzone2}\\ \end{tabular}& \begin{tabular}[c]{@{}l@{}}Mix-cont\\ext-based\end{tabular} & Infrastructureless & No & No & No & ASR & Analy-modekl  \\      
			
			\hline
			\begin{tabular}[c]{@{}l@{}}SocialSpots\\ \cite{socialconf,social} \end{tabular}&  \begin{tabular}[c]{@{}l@{}}Mix-zone-\\ based \end{tabular} & \begin{tabular}[c]{@{}l@{}}Infrastructure-based \end{tabular}   & No & No & No & ASS & \begin{tabular}[c]{@{}l@{}} Analy-model \\Simulation \end{tabular}  \\      
			
			\hline
			\begin{tabular}[c]{@{}l@{}}Mix-Context- \\Enhanced \cite{mixcontextenhanced}\end{tabular} & \begin{tabular}[c]{@{}l@{}}Mix-cont\\ext-based \end{tabular} & Infrastructureless & No & No & No & Statistics & Simulation  \\ 
			
			\hline
			\begin{tabular}[c]{@{}l@{}} {REP} \cite{rep} \\ \end{tabular} & \begin{tabular}[c]{@{}l@{}}Mix-cont\\ext-based \end{tabular} & Infrastructureless  & No & Yes & No & ASS  & Simulation   \\ 
			
			\hline \begin{tabular}[c]{@{}l@{}} CPN \cite{Cooperative} \\ \end{tabular}& \begin{tabular}[c]{@{}l@{}}Mix-cont\\ext-based \end{tabular} & Infrastructureless   & No & No & No & ASS  & \begin{tabular}[c]{@{}l@{}} Analy-model \\Simulation \end{tabular}  \\ 
			
			\hline
			\begin{tabular}[c]{@{}l@{}}SlotSwap \cite{sswapp,slotswap} \\ \end{tabular}& \begin{tabular}[c]{@{}l@{}}Mix-context-\\ based \end{tabular} & Infrastructureless & No & No & Yes & Entropy & Simulation \\  
			
			\hline
			\begin{tabular}[c]{@{}l@{}} DMLP \cite{dynamicmix,motivationbased}\\\end{tabular}& \begin{tabular}[c]{@{}l@{}}Mix-cont\\ ext-based \end{tabular} & \begin{tabular}[c]{@{}l@{}}Infrastructure-based \end{tabular}  & No & Yes & No & Entropy & Simulation \\ 
			
			\hline
			\begin{tabular}[c]{@{}l@{}}PCC \cite{condidates} \\ \end{tabular}& \begin{tabular}[c]{@{}l@{}}Mix-cont\\ ext-based \end{tabular} & Infrastructureless & No & No & No & \begin{tabular}[c]{@{}l@{}}ASS \\ ASR \end{tabular} & Simulation \\ 
			
			\hline
			
			\begin{tabular}[c]{@{}l@{}}SPCP \cite{spcp} \\ \end{tabular}& \begin{tabular}[c]{@{}l@{}}Mix-cont\\ ext-based \end{tabular} & Infrastructureless & No & No & No & \begin{tabular}[c]{@{}l@{}}ASS\\ASR \end{tabular} & Simulation \\ 
			
			\hline
			
			\begin{tabular}[c]{@{}l@{}}UPCS \cite{s2si,upcs}  \\ \end{tabular}&  \begin{tabular}[c]{@{}l@{}}Mix-zone- \\ based \end{tabular}   & \begin{tabular}[c]{@{}l@{}}Infrastructure-based \end{tabular}& Yes & No & Yes & Entropy & \begin{tabular}[c]{@{}l@{}} Analy-model \\ Simulation  \end{tabular}  \\ 
			\hline
			\begin{tabular}[c]{@{}l@{}}TAPCS \cite{tapcs}   \\ \end{tabular} & \begin{tabular}[c]{@{}l@{}}Mix-cont \\ ext-based \end{tabular} &  Infrastructureless & Yes & No & No & Entropy & \begin{tabular}[c]{@{}l@{}} Analy-model \\ Simulation  \end{tabular}  \\ 
			
			\hline
			\begin{tabular}[c]{@{}l@{}}VLPZ-based \\PCS \cite{vlpz, toward_efficient} \end{tabular}& \begin{tabular}[c]{@{}l@{}}Mix-zone- \\ based \end{tabular} & \begin{tabular}[c]{@{}l@{}}Infrastructure-based \end{tabular}& Yes & No & No & \begin{tabular}[c]{@{}l@{}} ASS \\ DA \end{tabular}  & \begin{tabular}[c]{@{}l@{}} Analy-model \\ Simulation  \end{tabular} \\ 
			\hline 
			\toprule
		\end{tabular} 
	}
	
	\caption{A summary of pseudonym changing strategies.}
	\label{tab:summary}
	
\end{table*}

\section {Comparison \& Discussion}
\label{sec:disscussion}

In this section, we discuss the effectiveness of pseudonym changing strategies. We based our analyses on a strong passive adversary model, consisting of an External Global Passive Adversary  and an Internal Local Passive Adversary composed of few attackers. A realistic study case of the assumed adversary model can be given as follows. Similarly to a mobile network, a VANET infrastructure is envisioned to be managed by vehicular system operators. It could exist for example a corrupt employee that works at an operator and has a full access to the infrastructure administration system. Since this employee is able to capture each event occurs in the VANET system, he can be considered as an external global passive adversary. Moreover, this employee can collude with some VANET's users (drivers) to help him in the tracking of their targets. These users can then be considered as an internal local passive adversary. In this discussion, we study the location privacy protection provided by the strategies against each part of the adversary model (internal or external) separately.  

By synchronizing pseudonym changing processes between vehicles, all strategies provide some level of protection against the syntactic linking of pseudonyms attack. This level of protection depends on the accuracy of the applied synchronization method and the number of vehicles that involved in this process.  After analyzing the strategies, we distinguished three used synchronization methods. We could sort them according to their effectiveness against the syntactic linking attack as follows:

\begin{enumerate}
	\item GPS-based synchronization.
	\item Infrastructure-based synchronization.
	\item Protocol-based synchronization. 
\end{enumerate}

We argue that the strategies based on the GPS synchronization method are the most effective against the syntactic linking attack, because they involve all vehicles. In addition, using the GPS signals is considered among the most accurate time synchronization methods. Indeed, the accuracy of GPS time signals is at the level of a few nanoseconds \cite{GPS}. However, GPS signals could be affected due to the bad weather and cannot be received in the tunnels, underground passages, or near tall buildings \cite{gspaffected}.  The infrastructure-based method is the second more effective synchronization method, because it involves a limited number of vehicles i.e. only vehicles exist inside the zone controlled by the infrastructure. As the external adversary cannot control the zone, it seems that all pseudonyms changes occur simultaneously. The protocol-based synchronization method is less effective than the others, because it involves a limited number of vehicles and it is not guaranteed if all the concerned vehicles will change their pseudonyms or not. 

Regards the protection against the semantic linking attack, we noticed that just few strategies can provide protection against this attack. These strategies are based on hiding safety messages information from the adversary for some period of time. To achieve this, two techniques have been proposed: the encryption and the radio silence. However, each of these two techniques has drawbacks. On the one hand, the encryption of safety messages is ineffective against the internal passive adversary. Indeed, as this adversary has the privileges (credentials) to decrypt safety messages, it can then provide a clue to the external adversary to read the contents the safety messages. In addition, encrypting safety messages may not meet with the requirements of VANETs and may introduce a communication overhead due to the messages used to share the encryption keys \cite{cmix}. On the other hand, because no safety message is broadcasting during the radio silence period, the use of this technique is challenging. Indeed, the radio silence may affect  the VANETs' safety-related applications if it is not used properly \cite{safety}.  However, compared to the encryption technique, the radio silence provides an effective protection against both internal and external passive adversaries, since no safety message is provided to the adversary during the radio silence. Therefore, we argue that the strategies that based on the radio silence are more effective against the semantic linking attack compared to the strategies that based on the encryption. 

Besides this, the technique of the exchanging of pseudonyms was used in SlotSwap \cite{slotswap}. This technique is useful, because it limits the number of pseudonyms used by vehicles, which has positive impacts on network performances and vehicles' storage. However, exchanging of pseudonyms without informing the authorities leads to losing the accountability/liability, which is one of the primary security requirements for VANETs.  In addition, the overhead may be introduced, because of the huge number of messages that can be used in the exchange of pseudonyms.


\begin{table*}[!ht]
	\centering
	\resizebox{\textwidth}{!}{
		\begin{tabular}{l l l l l l l l l l l}
			
			\hline 
			\toprule
			
			\multirow{3}{2.5 cm}{{\textbf{Strategy}}} & 
			\textbf{Synchronization} & 
			\textbf{Semantic}     & 
			\textbf{Syntactic}    &
			\textbf{Semantic}  
			
			\\ &\textbf{Method}& \textbf{Protection} & \textbf{Protection}  & \textbf{Protection}  
			\\ &               & \textbf{Technique} & \textbf{Against}  & \textbf{Against}  \\
			\hline 
			\toprule
			
			CMIX              &  Infrastructure-based  & Encryption  & External only & External only    \\
			
			\hline
			CARAVAN/AMOEBA	    &  Protocol-based       & Radio silence & Both        & Both            \\
			
			\hline
			Mix-Context         &  Protocol-based       & None          & Both         & None           \\ 
			
			\hline
			SLOW               &  Protocol-based       & Radio silence & Both         & Both           \\
			
			\hline
			DLP                &  Protocol-Based       & None          & Both         & None           \\      
			
			\hline
			SocialSpots         &  Infrastructure-based & None          & Both         & None           \\      
			
			\hline
			Mix-ContextEnhanced & Protocol-based        & None          & Both         & None           \\ 
			
			\hline
			REP                 &  Protocol-based       & Encryption    & External only& External only  \\ 
			\hline
			CPN                 &  Protocol-based       & None          & Both         & None           \\ 
			
			\hline
			SlotSwap            &  GPS-based            & None          & Both         & None           \\  
			
			\hline
			DMLP                &  Protocol-based       & Encryption    & External only& External only  \\ 
			
			\hline
			PCC                 &  Protocol-based       & None          & Both         & None           \\ 
			
			\hline
			SPCP                &  Protocol-based       & None          & Both         & None           \\ 
			\hline
			UPCS & Infrastructure-based & Radio silence & Both & Both \\ 
			\hline
			TAPCS & Protocol-based & Radio silence & Both & Both  \\ 
			\hline
			VLPZ-based PCS & Infrastructure-based & Radio silence & Both & Both  \\ 
			\hline 
			\toprule
		\end{tabular} 
	}
	
	\caption{A review of pseudonym changing strategies.}
	\label{tab:reviewpseudonymstrategy}
	
\end{table*}

To summarize, Table~\ref{tab:reviewpseudonymstrategy} reviews pseudonym changing strategies based on the following parameters:

\begin{itemize}
	
	\item \textbf{Synchronization method:} indicates which method is used to synchronize the pseudonyms changes between vehicles.
	\item \textbf{Semantic protection technique:} indicates which technique is used to temporary hide the safety message's content from the adversary.
	\item \textbf{Syntactic protection against:} indicates against which type of adversary (internal/external), the syntactic protection is provided. "Both" stands for both of the two types.
	\item \textbf{Semantic protection against}: indicates against which type of adversary (internal/external), the semantic protection is provided. "Both" stands for both of the two types.
\end{itemize}

\begin{table*}[!ht]
	\centering
	\resizebox{\textwidth}{!}{
		\begin{tabular}[]{ l|l l|l l l }
			
			\hline  & 
			\multicolumn{2}{c|}{ {Privacy Protection}}& \multicolumn{3}{c}{Costs} \\ 
			\hline \multirow{3}{2.5 cm}{{\textbf{Strategy}}}  & \textbf{Syntactic} & \textbf{Semantic}   & \textbf{Impacts on}  & \textbf{Overhead} & \textbf{Accountability}  \\ 
			&               \textbf{Linking } & \textbf{Linking}  & \textbf{Safety}&  & \textbf{loss}  \\
			&             \textbf{Protection} & \textbf{Protection}  &  &  &   \\
			
			\hline
			CMIX                  & +    & +  & No & Yes & No \\ 
			\hline
			CARAVAN/AMOEBA        & ++   & ++ & - - - & No  & No \\ 
			\hline
			Mix-Context           & ++   & No & No & No  & No \\ 
			\hline
			SLOW                  & ++   & ++ & - -  & No  & No \\
			\hline
			DLP                   & ++   & No & No & No  & No \\
			\hline
			SocialSpots           & +++  & No & No & No  & No \\
			\hline
			Mix-Context-Enhanced  & ++   & No & No & No  & No \\ 
			\hline
			REP                   & +    & +  & No & Yes & No \\ 
			\hline
			CPN                   & ++   & No & No & No  & No \\
			\hline
			SlotSwap              & ++++ & No & No & Yes & Yes \\ 
			\hline
			DMLP                  & +    & +  & No & Yes & No \\
			\hline
			PCC                   & ++   & No & No & Yes & No \\
			\hline 
			SPCP                  & ++   & No & No & Yes & No \\
			\hline
			UPCS                  & +++  & ++ & -  & No  & No \\
			\hline
			TAPCS                 & ++   & ++ & No & No  & No \\
			\hline
			VLPZ-based PCS        & +++  & ++ & No & No  & No \\
			\hline 
		\end{tabular} 
	}
	\caption{A comparison between pseudonym changing strategies.}
	\label{tab:one}
\end{table*}
In Table~\ref{tab:one}, we compare pseudonym changing strategies. This comparison is based on two axes:

\begin{enumerate}
	\item \textbf{The level of the pseudonyms linking prevention:} this is measured by the ability of the strategy to prevent the syntactic and the semantic linking attacks of pseudonyms. We define four levels for \textit{the syntactic linking protection} based on "Synchronization method" and "Syntactic protection against" parameters (see Table~\ref{tab:two}). We also define two level for \textit{the semantic linking protection} based on  "Semantic protection against" parameter (see Table~\ref{tab:three}).

	\item \textbf{The costs involved in the changing of the pseudonym:} which comprise:
	\begin{itemize}
		\item \textbf{Impacts on road safety:} indicate whether the strategy has negative impacts on road safety  or no. Three levels of this parameter are defined. 
		\item  \textbf{Overhead:} this is expressed in terms of the additional messages that can be used by the strategy to obtain the encryption keys or exchange the pseudonyms for example.
		\item  \textbf{Accountability Loss:} indicates if the strategy leads to the loss of accountability or not.
	\end{itemize}

\end{enumerate}

\begin{table*}[!ht]
	\centering

	\begin{tabular}[lr]{l l l}
		\toprule
		Syntactic Protection & Syntactic Protection  & Synchronization Method\\
		Level& Against & \\
		\toprule
		+                      & External only                                 &   {          *          }                                       \\
		\hline{++}                         & Both                                          & Protocol-Based                           \\ 
		\hline{+++}                        & Both                                          & Infrastructure-based                     \\ 
		\hline{++++}                       & Both                                          & GPS-Based                                \\ 
		\toprule 
	\end{tabular}

	\caption{Syntactic protection levels.}	
	\label{tab:two}
\end{table*}

\begin{table*}[!ht]
	\centering
	
	\begin{tabular}[lr]{c c}
		\hline{Semantic Protection level} & {Semantic protection against}                 \\
		\hline{+}                          & External only                                 \\
		\hline{++}                         & Both                                          \\ 
		\hline 
	\end{tabular} 
	
	\caption{Semantic protection levels.}
	\label{tab:three}
\end{table*}

\section{Open research issues}

Several parameters involves to build an effective pseudonym changing strategy. For example, the simultaneous pseudonyms change of a large number of vehicles is required. In addition, the use of the radio silence seems to be inevitable towards an effective strategy. However, although considerable efforts have been made, several open issues are still needing more attention to achieve the aimed strategy. Some of these issues are discussed in the following section.

\subsection{Impact on road safety}

As discussed in Section~\ref{sec:disscussion}, the radio silence technique has proved its effectiveness to protect against both external and internal passive adversaries. However, using this technique has negative impacts on safety-related applications \cite{safety2}. In \cite{safety}, Lef\`{e}vre et al. studied the impact of the radio-silence based pseudonym changing strategies on Intersection Collision Avoidance (ICA) Systems. Their results highlight a positive correlation between the radio silence duration and the impact on road safety. For example, their simulation results show, in case that the strategy of pseudonym changing proposed by the SAE J2735 standard \cite{sae} is used, the radio silence duration should be shorter that two seconds to ensure the well function of the considered ICA application. For this reason, the authors proposed an adaptive pseudonym changing strategy that compromises between privacy and safety. They conclude that the requirements of both safety-related applications and  pseudonym changing strategies should be taken into account in the design of each of them.

In the last few years, several researchers start focusing on this issue. The existing works can be divided into two categories. The works of the first category ( e.g. \cite{ontheevaluation}, \cite{caps}, \cite{poster}, \cite{context}, \cite{tapcs}) suggest that vehicles continuously monitor its neighborhood to find a good opportunity when the radio silence can be used. In the other hand, the works of the second category suggest executing a radio-silence based strategy in places where the impact on road safety is low  or null such as signalized intersections \cite{s2si} \cite{upcs} and widespread roadside infrastructures \cite{vlpz,toward_efficient} (e.g. gas stations, electric vehicles charging stations, and toll booths).

\subsection{Non cooperative behavior}

The cooperation between vehicles is a key factor on a successful pseudonym changing strategy. However, due to the costs that involve in changing the pseudonym such as the overhead, and the cost of changing and managing the pseudonym, vehicles may not want to cooperate with other vehicles. The non-cooperative behavior in pseudonym changing strategy is first studied by Freudiger et al. in \cite{noncooperative}. The authors proposed a game-theoretic model that takes into account the gained payoff and the cost generated by each vehicle while executing the strategy. They demonstrated the existence of the Nash equilibria in both static and dynamic forms of the game within complete  and incomplete information. Based on the results of their analysis and simulations, they proposed PseudoGame protocols for an optimal pseudonym changing. The cooperation behavior is studied in \cite{motion} using the auction game model as well.

However, when we analyzed strategies we found that just few strategies take into account the non cooperative behavior. Indeed, the authors in \cite{social} relied on a simplified game-theoretic to demonstrate the feasibility of the proposed strategy assuming that all vehicles are rational. The authors in \cite{reputationbased,motivationbased}, studied the selfish behavior in the DMLP pseudonym changing strategy. They proposed a reputation mechanism to stimulate selfish vehicles to cooperate by changing their pseudonyms in a DMLP. The reputation value of a vehicle is increased each time it cooperates at other vehicle's DMLP. The increase of reputation is computed based on the number of cooperating vehicles at the $i^{th}$ DMLP. The accumulated reputation strength of a vehicle until the $i^{th}$ DMLP is used as a credit when a vehicle requests to create its own DMLP. Indeed, if the  accumulated reputation strength $R^{i}$ is greater or equal to a certain threshold strength of reputation $\varepsilon$, all vehicle that receive the COMMAND message will change their pseudonyms. Else, the decision of a vehicle to cooperate or not depends to its current reputation $R^{j}$, its current location privacy level, the remaining lifetime of the current pseudonym (T). Indeed (i) if $R^{j}$ is less than $\varepsilon$, a vehicle has to cooperate to increase its reputation value, (ii) if the location privacy level is the less then recommended level $\gamma$ the vehicle has to change its pseudonym as well, and (iii) if the remaining life of the current pseudonym is close to (T), the vehicle will change its pseudonym. Otherwise, the vehicle keep its pseudonym. 

The authors in \cite{toward_efficient} also used a reputation based mechanism to motivate rational vehicles to enter the VLPZ. Indeed, as the level of privacy protection mainly depends on the number of vehicles inside the VLPZ at same time, the VLPZ always tries to increase its occupancy. A vehicle is allowed to execute the strategy only if its reputation value is above or equal to a certain threshold ($\omega$) or it has not already refused an invitation form a VLPZ. Indeed, the VLPZ occasionally sends invitations to vehicles to motivate them to enter to it. If a vehicle accepts to enter, its reputation value will be increased. However, if a vehicle refuses, its reputation value will be decreased. The following formulas gives the reputation value  $\mathbb{R}_{i}^{j}$ of a given vehicle (\textit{v$_i$}) after the $j^{th}$  invitation.

\[
\mathbb{R}_{i}^{j}= 
\left \{
\begin{tabular}{ll}
$\mathbb{R}_{i}^{j-1} + |AS|_{t_{j}}$                                                        & if v cooperates  \\
&      \\
$ \mathbb{R}_{i}^{j-1} - |AS|_{t_{j}}$                                                       & if v defeats and $\mathbb{R}_{i}^{j-1} \geqslant  |AS|_{t_{j}}$ \\
&      \\
0                                                       &  if v defeats and $\mathbb{R}_{i}^{j-1} < |AS|_{t_{j}}$ \\

\end{tabular}
\right. 
\]

Where $\mathbb{R}_{i}^{j-1}$ is the old reputation value of \textit{v$_i$}. The reputation value of the vehicle increases as much as it cooperates. The increase or the decrease of the reputation value depends on the VLPZ occupancy at t$_{j}$, where the time t$_{j}$ is depended on the decision of \textit{v$_i$}. 

\subsection{Evaluation metrics and techniques}
Due to the abstract nature of privacy concept, quantifying the privacy protection level is difficult task to perform. A variety of metrics (e.g. \cite{privacymetrics}\cite{eval1}\cite{eval3}\cite{eval4}\cite{eval5}) and techniques (e.g.\cite{eval2}\cite{eval6}\cite{eval7}\cite{eval8}\cite{eval9})  have been proposed to evaluate the achieved level of location privacy protection in VANETs. However, finding an unified framework and a comprehensible metric to evaluate and compare pseudonym changing strategies is not yet achieved. 

Three initiatives have recently been emerged to propose a simulation framework to evaluate the privacy level in VANETs. Tomandl et al. \cite{eval7} proposed VANETsim an event-driven simulation framework that allows implementing a set of pseudonym changing strategies in an abstract way i.e. without considering the wireless medium characteristics and communication protocols used in VANETs. In \cite {eval6}, Eckhoff  et al. described a generic design of Veins simulation framework extension devoted to evaluate the pseudonym changing strategies. This extension consists of three main blocks (i) the attacker model block, (ii) the metrics block, and (iii) the scenarios block. This extension is concertized by Emara in \cite {prext}. Indeed, the authors proposed PREXT (Privacy Extension for Veins VANET Simulator). PREXT adopts the attacker model proposed in \cite{eval1}, uses most popular privacy metrics and implements seven pseudonym changing strategies.

From a macroscopic view, these efforts seem encouraging. However, with more focus, we believe that more efforts should to be done to evaluate the proposed privacy mechanisms. In particular, the used privacy metrics are general and do not take the context of VANETs into account. We think that the difficulty of the problem is mainly related to the definition of privacy in the context of VANETs.

\section{Conclusion}

The pseudonym changing strategy of is an important building block of the pseudonym changing approach. All efforts should be made in order to achieve the aimed strategy before the real-word deployment of the pseudonym changing mechanism. In this survey paper, we surveyed relevant pseudonym changing strategies for VANETs and classified them into two categories. We also identified the strengths and costs generated by the presented strategies. Finally, we highlighted some challenges on the pseudonym changing strategy issue.



\bibliographystyle{IEEEtran}
\bibliography{IEEEabrv,survey_refs}

\end{document}